\documentclass[fleqn,usenatbib,useAMS]{mnras}
\usepackage{newtxtext,newtxmath}

\usepackage{graphicx}	
\usepackage{amsmath}	
\usepackage{multicol}        
\usepackage{bm}		
\usepackage{pdflscape}	

\usepackage[normalem]{ulem} 
\usepackage[dvipsnames]{xcolor}
\usepackage{soul}

\usepackage{ulem} 

\newcommand{\Ms}{ \mathrm{M}_\odot }
\newcommand{\kms}{ \mathrm{kms}^{-1}}

\title[The Timeless Timing Argument]{The Timeless Timing Argument and the Total Mass of the Local Group}

\author[Sawala et al.]{Till Sawala$^{1,2}$\thanks{Email: till.sawala@helsinki.fi}, Jorge Peñarrubia$^{3}$, Shihong Liao$^{1}$, and Peter~H.~Johansson$^{1}$ \\
$^{1}$Department of Physics, University of Helsinki, Gustaf H\"allstr\"omin katu 2, FI-00014 Helsinki, Finland \\
$^{2}$Institute for Computational Cosmology, Durham University, South Road, Durham DH1 3LE, United Kingdom\\
$^{3}$Institute for Astronomy, University of Edinburgh, Royal Observatory, Blackford Hill, Edinburgh EH9 3HJ,  United Kingdom\\
}

\date{Accepted XXX. Received YYY; in original form ZZZ}
\pubyear{2022}

\begin{document}
\label{firstpage}
\pagerange{\pageref{firstpage}--\pageref{lastpage}}
\maketitle

\begin{abstract}
The Timing Argument connects the motion of a two-body system to its mass in an expanding Universe with a finite age, under the assumption that it has evolved on a self-gravitating orbit. It is commonly applied to the present-day Milky Way-M31 system in order to infer its unknown mass from the measured kinematics. We use a set of Local Group analogues from the {\sc Uchuu} simulation to investigate the Timing Argument over cosmic time. We find that the median inferred mass remains almost constant over the past 12 Gyr, even while the haloes themselves grew in mass by more than an order of magnitude. By contrast, we find a closer, and nearly time-invariant agreement between the Timing Argument value and the mass within a sphere of radius equal to the MW-M31 separation, and we identify this as the total mass of the system. We conclude that the comparatively close present-day agreement between the Timing Argument and the sum of the halo masses reflects no underlying relation, but merely echoes the fact that the MW and M31 now contain most (but not all) of the mass of the Local Group system.
\end{abstract}

\begin{keywords}
(galaxies:) Local Group, Galaxy: kinematics and dynamics, cosmology: theory, dark matter, methods: numerical
\end{keywords}



\section{Introduction}
The standard structure and galaxy formation theory \citep{White-1987, White-1991} predicts that most of the mass of our own Milky Way (MW), of the neighbouring Andromeda galaxy (M31), and of the Local Group (LG), is contained in dark matter, which is presently detectable only through its gravitational effect on the visible components of galaxies and on the orbits of the galaxies themselves. 

An elegant approach to infer the total mass of the Local Group is presented in the `Timing Argument' \citep{Kahn-1959, Lynden-Bell-1981}, which relates it to the observed kinematics of the LG and the age of the Universe. This Argument assumes that the present MW - M31 system is on its first approach in a two-body orbit which expanded after the Big Bang before collapsing under its own gravity. Initially applied under the assumption of a purely radial orbit \citep{Li-2008, vanderMarel-2008, vanderMarel-2012}, the Timing Argument has since been extended to include eccentricity \citep{Li-2008}, the effects of dark energy \citep{Partridge-2013,Benisty-2023, McLeod-2020} and modified gravity \citep{Benisty-2023b, McLeod-2020}, cosmic bias \citep{vanderMarel-2012}, or the recoil velocity of the MW with respect to the LMC \citep{Benisty-2022, Chamberlain-2023}. It has been invoked to exclude a past encounter of the MW and M31 \citep{Benisty-2021}, and has also been used to model the dynamics of galaxies in the Local Volume \citep{Lynden-Bell-1981, Penarrubia-2014, Penarrubia-2015}. As our focus is on the fundamental relation between the Timing Argument and the mass of the Local Group, rather than on a precise mass estimate, we only consider the Keplerian case, but our findings should equally apply to more complex models.

Comparisons between results from the Timing Argument and other mass estimates or results from numerical simulations \citep[e.g.][]{Li-2008, Guo-2011, Lemos-2021, Hartl-2022, Sawala-2023} generally equate the former with the sum of the two halo masses, each defined as the virial mass, the bound mass, or via a density threshold. The Timing Argument itself does not invoke any specific mass definition, but instead a `total mass' of the system, a rather ill-defined quantity in cosmology.

In this work, we aim to identify the mass associated with the Timing Argument in the Local Group and, closely related, extend the previous analyses beyond $z=0$. This is particularly significant because while the implicit assumption of constant halo masses evidently breaks down at early times, the Timing Argument itself is time-independent. Unlike a definition based on halo masses, the mass associated with the Timing Argument should be {\it timeless}.

\section{The Timing Argument} \label{sec:TA}
The Timing Argument derives the mass, $M$, and other orbital variables from the measured separation, $r$, radial velocity, $v_r$, and transverse velocity, $v_t$, via the following set of equations \citep[e.g.][and references therein]{vanderMarel-2008}:

\begin{eqnarray}\label{eqn:TA-beginning}
    r &=& a \left(1 - e \: \mathrm{cos} \: \eta\right), \\
    t &=& \left(\frac{a^3}{GM}\right)^{1/2} \left(\eta - e \: \mathrm{sin} \: \eta\right), \\
    v_r &=& \left(\frac{GM}{a}\right)^{1/2} \frac{e \: \mathrm{sin} \: \eta} {1 - e \: \mathrm{cos} \: \eta}, \\
    v_t &=& \left(\frac{GM}{a}\right)^{1/2} \frac{(1 - e^2)^{1/2}}{1 - e \: \mathrm{cos} \: \eta},\label{eqn:TA-end}
\end{eqnarray}
where $e$ is the orbital eccentricity, $a$ is the semi-major axis, and $\eta$ is the orbital phase at (the known) time $t$ since the Big Bang. For a purely radial orbit, $v_t = 0, e = 1$, in which case the problem reduces to the three equations already given by \cite{Lynden-Bell-1981}.

Conceptually, the initial conditions of the orbit are set at the time of the Big Bang as $r(t=0)=0$ and $v_r(t=0)>0$, given by the Universal expansion. It is worth noting, however, that an orbit with non-zero angular momentum is not fully consistent with the Big Bang picture, and requires the action of external tidal forces caused by the evolving inhomogeneous matter distribution.

A direct numerical solution to Equations~\ref{eqn:TA-beginning}--\ref{eqn:TA-end} requires finely adjusted initial values, which proves impractical when the variables are not constrained to the observations at $z=0$. We therefore split the equations, obtaining a single equation with $\eta$ as the only unknown:

\begin{equation}\label{eqn:TA-split}
\frac{t \: v_r}{r} =
\frac{ \mathrm{sin} \: \eta \left(\eta \left(\left(\frac{v_t}{v_r}\right)^2 \mathrm{sin^2} \: \eta + 1\right)^{1/2} - \mathrm{sin} \: \eta\right)}
{\left(\mathrm{cos} \: \eta - \left(\left(\frac{v_t}{v_r}\right)^2 \mathrm{sin^2} \: \eta + 1\right)^{1/2}\right)^2}.
\end{equation}

To understand the behaviour of Equation~\ref{eqn:TA-split}, we note that, because $r$ and $t$ are both positive, both sides of the equation take the sign of $v_r$. For $v_r < 0$ (indicating the approach of two galaxies), the RHS can take values in the interval $(-\infty, 0)$, while for $v_r > 0$ (indicating that the galaxies are moving away from each other), it can take values in the interval $(0, 3/4)$. Orbits with $v_r > \frac{3}{4} \frac{r}{t}$ are unbound. For $\eta \in (0, \pi)$, the RHS is always positive, and for a given $t v_r/r$, Equation~\ref{eqn:TA-split} has at most two solutions. For $\eta \in (\pi, 2\pi)$, the RHS is negative, and for a given $t v_r/r$, the equation has at most one solution.

When tracing the Local Group analogue systems backwards in time, we begin with $v_{r,0} < 0$ in each case, resulting in a unique solution. At times when $v_r$ becomes positive and Equation~\ref{eqn:TA-split} admits two solutions, we choose the one that causes minimal discontinuities.

After solving Equation~\ref{eqn:TA-split}, we obtain values for $e$, $a$ and $M$ via:

\begin{eqnarray}\label{eqn:TA-analytical}
e &=& \left(\left(\frac{v_t}{v_r}\right)^2 \mathrm{sin^2} \: \eta + 1 \right)^{-1/2}, \\
a &=&  \frac{r}{1 - e \ \mathrm{cos} \: \eta} ,\\
M &=& \frac{a^3}{G t^2} \left(\eta - e \ \mathrm{sin} \: \eta \right)^2.
\end{eqnarray}

The mass, $M$, semi-major axis, $a$, and eccentricity, $e$, are all constants of the motion and as such should be constant over time. In Section~\ref{sec:time-evolution}, we examine to which extent this is true as we measure them for real orbits of Local Group analogue systems that form and evolve in a cosmological simulation.

When we determine the mass of the Local Group from the kinematics using Equations~\ref{eqn:TA-beginning}--\ref{eqn:TA-end}, we refer to it as $\mathrm{M_{TA}}$ to distinguish it from other mass definitions which we introduce in Section~\ref{sec:simulation}. Throughout this paper, we express masses in physical $\mathrm{M_\odot}$ and velocities in physical $\kms$, i.e. radial velocities include both the peculiar velocity and the contribution from the Hubble expansion. Distances are expressed as either physical kpc or comoving $c$kpc. To identify quantities measured at $z=0$, we use a $0$-subscript.

\begin{figure*}
    \centering
    \includegraphics[width=7.in]{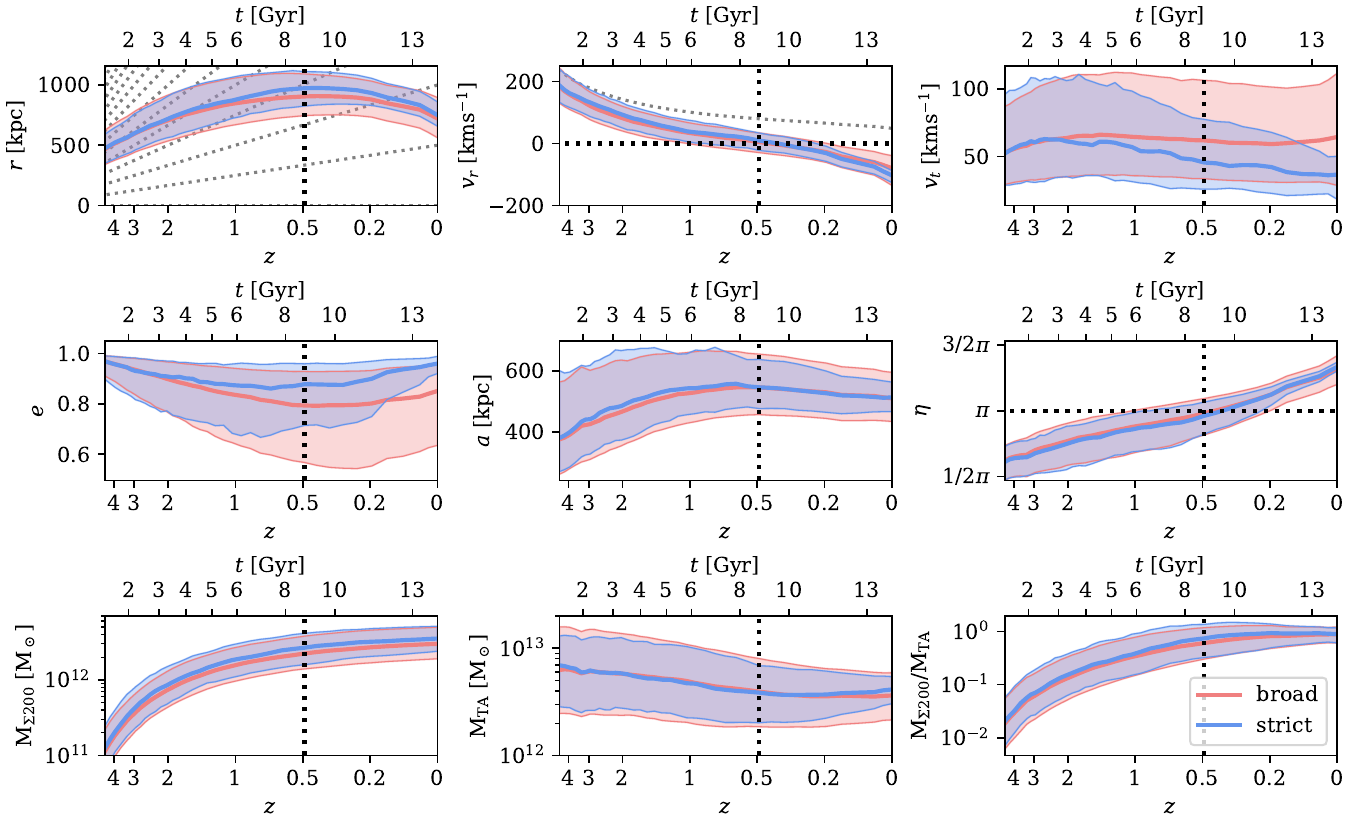}
    \vspace{-.1cm}
    \caption{Evolution of the three kinematic variables measured in the simulation, $r$, $v_r$ and $v_t$ (top row, from left to right), the variables of the Timing Argument, $e$, $a$ and $\eta$ (middle row, from left to right), the Timing Argument mass estimate, $\mathrm{M_{\Sigma 200}}$, $\mathrm{M_{TA}}$, and the ratio $\mathrm{M_{\Sigma 200}} / \mathrm{M_{TA}}$ (bottom row, from left to right). On all panels, blue lines show the median evolution of `strict' LG analogues, red lines show the median for `broad' LG analogues, shaded regions show percentiles equivalent to $\pm 1 \sigma$. Vertical dotted lines indicate the median time of maximum expansion. On the top left panel, grey dashed lines indicate the expansion of the Universe. On the top centre panel, the horizontal line indicates $v_r=0$, the grey dotted line indicates the Hubble flow.
    }
    \label{fig:evolution-Timing-Argument}
\end{figure*}

\section{Local Group analogues} \label{sec:simulation}
Our analysis is based on the {\sc Mini-Uchuu} simulation \citep{Ishiyama-2021}, containing several thousand LG analogues with enough resolution to resolve MW or M31 mass haloes and to trace their main progenitors over a large fraction of cosmic time. Crucially for our work, both merger trees and full particle data for all snapshots are publicly available. The {\sc Uchuu} simulations use cosmological parameters corresponding to the \cite{Planck-2020} results, namely, $\Omega_0=0.3089$, $\Omega_{\rm b}=0.0486$, $\Omega_{\Lambda} =0.6911$, $h=0.6774$, $n_s=0.9667$, and $\sigma_8=0.8159$. The simulation outputs have been processed using the {\sc Rockstar} phase-space structure finder \citep{Behroozi-2013a}.

We define the mass of each halo as $\mathrm{M_{200}}$, i.e. the mass enclosed within a sphere whose density is $200\ \times$ the critical density. Throughout this paper, we refer to the sum of the values of $\mathrm{M_{200}}$ for the MW and M31 analogues as $\mathrm{M_{\Sigma 200}}$ but note again that, although common, from a dynamical perspective, it is a rather arbitrary definition.

We select LG analogues based on their present day kinematics. We require two haloes with an $\mathrm{M_{200,0}}$ mass in the range $[0.5 - 5] \times 10^{12} \mathrm{M_\odot}$, and a mass ratio no more than 5 between the two haloes. Following \cite{Sawala-2023}, we also require that the LG analogues form `true' pairs at $z=0$, that is, their mutual interactions dominate over those with any third haloes.

Our goal is not only to understand the Timing Argument as it pertains to our Local Group, but also investigate it for analogues more broadly. For this reason, we select pairs with two sets of kinematic criteria. We identify a set of 425 `strict' LG analogues with $r_0 = 770 \pm 150$~kpc, $v_{r,0} = -110 \pm 30$~kms$^{-1}$ and $v_{t,0} < 60$~kms$^{-1}$, and a larger set of 4902 `broad' LG analogues with $r_0 = 770 \pm 300$~kpc, $v_{r,0} = -110 \pm 90$~kms$^{-1}$ and $v_{t,0} < 150$~kms$^{-1}$.

For each system, we also measure the mass enclosed in spheres centred on the midpoint of the two haloes using the simulation particles, irrespective of whether or not they are within $r_{200}$ of either halo. We obtain sequences of mass measurements by varying the radius of the sphere. We refer to these masses as $\mathrm{M}(< R)$.

In order to trace their evolution, haloes are linked using merger trees, which have been constructed using the {\sc Consistent Tree} algorithm \citep{Behroozi-2013b}. We follow the most massive progenitors, and include only those pairs where both haloes can be traced back to $z=4.63$, which includes $96 \%$ of haloes and $93\%$ of pairs.

We apply a Savitzky-Golay filter \citep{Savitzky-1964} to the time series of all properties measured directly in the simulation. This prevents discontinuities in case halo positions and velocities change discontinuously during mergers, whilst ensuring that gradual changes to the kinematics caused by mergers and interactions are preserved. We measure the halo masses $\mathrm{M_{\sum 200}}$ and $\mathrm{M}(<R)$ at each output, and also solve the Timing Argument equations at every snapshot.

\section{Time evolution}\label{sec:time-evolution}
In Figure~\ref{fig:evolution-Timing-Argument}, we show the time evolution of the LG analogues. On each panel, thick lines showing the median value are bracketed by areas indicating the equivalents of $\pm 1 \sigma$ scatter. Blue and red colours represent the `strict' and `broad' samples, respectively.

The top row shows the three variables of the LG analogues measured in the simulation: separation, $r$, radial velocity, $v_r$, and transverse velocity, $v_t$. Initially, the LG analogues expand with the expansion of the Universe, reaching apocentre with a median maximum of $r \sim 1000$ kpc at $z \sim 0.5$ ($t\sim 9$~Gyr), after which the median radial velocity transitions from positive to negative.

The median transverse velocity, $v_t$, being unaffected by the Hubble expansion, shows much less evolution. We see a late-time decrease in $v_t$ between for the `strict' set only, explained by the fact that $v_t$ for each individual system is generated by random torques, and LG analogues in the `strict' set are selected to have low $v_t$ at $z=0$. In both sets, we also see a growth of $v_t$ before $z=3$, which we attribute to tidal torquing, analogous to the effect for proto-galaxies \citep{Hoyle-1951,Peebles-1969,Doroshkevich-1970,White-1984}.

The fact that the distributions converge towards higher redshifts also indicate that differences between the `strict' and `broad' sets are primarily caused by short-term, random variations. It is another manifestation of the irreducible uncertainty in inferring fundamental LG properties from present-day measurements \citep{Sawala-2023}.

In the middle row of Figure~\ref{fig:evolution-Timing-Argument}, we show three components of the orbital solution that constitutes the Timing Argument: the eccentricity, $e$, semi-major axis, $a$, and orbital phase, $\eta$. The first two show only limited evolution. Both sets of LG orbits are highly eccentric, with median values of $e_0=0.95$ for the `strict', and $e_0=0.85$ for the broad set. The median eccentricities remain above $e\gtrsim 0.8$ throughout the evolution. The median inferred semi-major axis is $a_0 \sim 500$~kpc in both cases, and $a$ remains between $\sim 400$ and $\sim 550$~kpc throughout the evolution. As expected, the orbital phase increases continuously, to $\eta_0 \sim 1.3 \times \pi$. Although not perfect, the median LG analogue is a reasonably close approximation to a system evolving along a two-body orbit.

\begin{figure*}
    \centering
   \includegraphics[width=7.in]
    {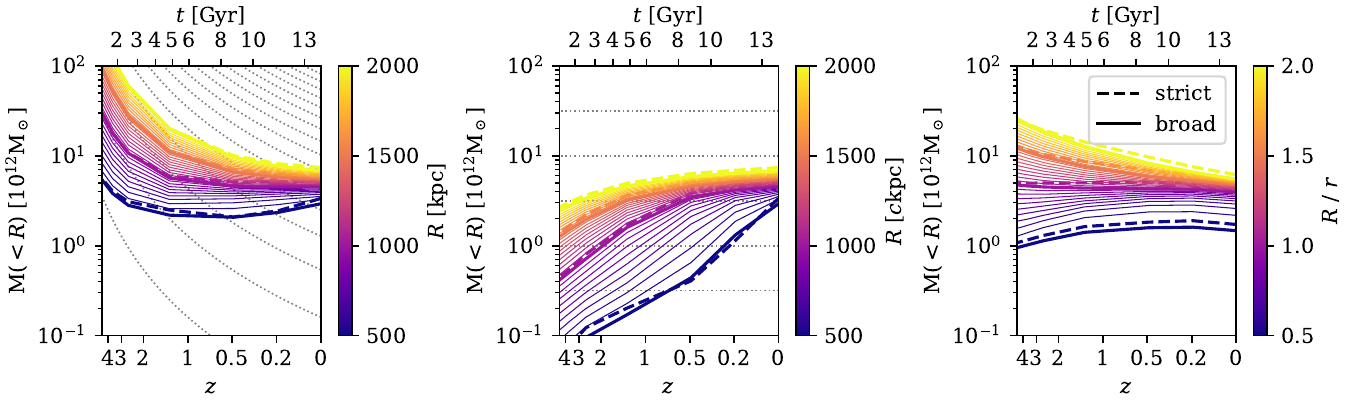}
    \caption{Evolution of the median mass enclosed within spheres of fixed physical radii (left), fixed comoving radii (middle) or multiples of the separation (right). On the left and the middle panel, thin dotted lines show the evolution assuming a density that evolves with the cosmic expansion. For any fixed physical or fixed comoving radius in the range $500 - 2000$~kpc, or $500 - 2000$~$c$kpc, the enclosed mass changes over time. However, the mass enclosed within a sphere of radius equal to the separation, $R= 1.0 \times r$, is nearly constant. There is no significant difference between the `broad' or `strict' sets.}
    \vspace{-.2cm}
    \label{fig:evolution-mass-estimators}
\end{figure*}

\begin{figure}
\centering
   \includegraphics[width=3.3in]{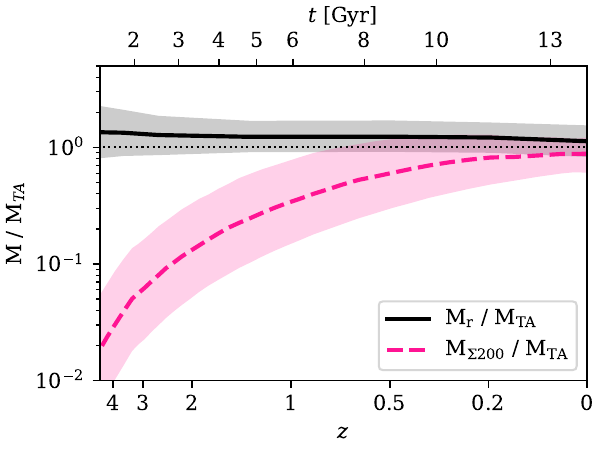}
    \caption{Evolution of the median ratio between the mass enclosed within a sphere of radius equal to the separation, $\mathrm{M_r}$, and the mass according to the Timing Argument, $\mathrm{M_{TA}}$, (black solid line), and of the median ratio between the sum of the halo masses, $\mathrm{M_{\Sigma 200}}$, and the mass according to the Timing Argument, $\mathrm{M_{TA}}$ (pink dashed line). Results shown are for the `broad' set but the `strict' set is nearly identical. Shaded regions indicate $\pm 1 \sigma$ scatter.}
    \label{fig:evolution-relative}
        \vspace{-.2cm}
\end{figure}

\begin{figure*}
   \includegraphics[width=2.25in]{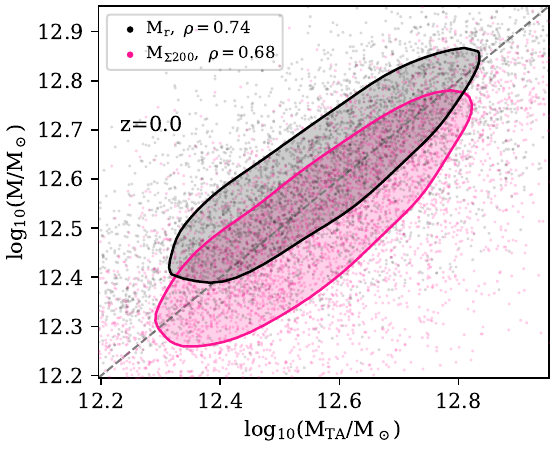} 
   \includegraphics[width=2.25in]{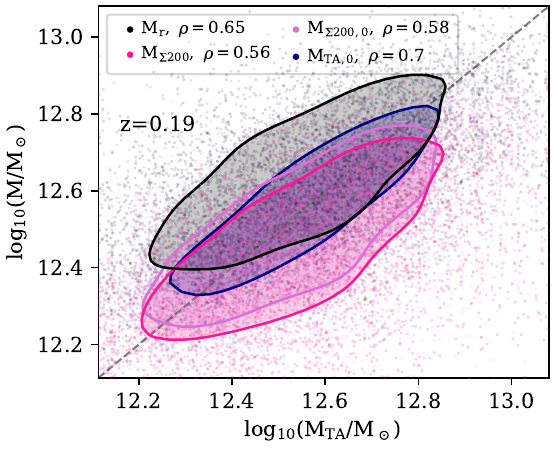} 
   \includegraphics[width=2.25in]{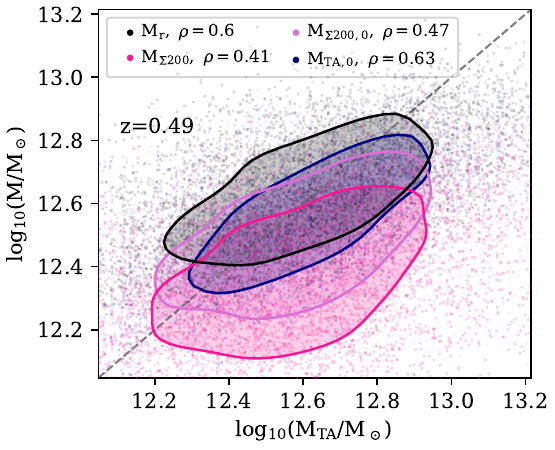} \\
   \includegraphics[width=2.25in]{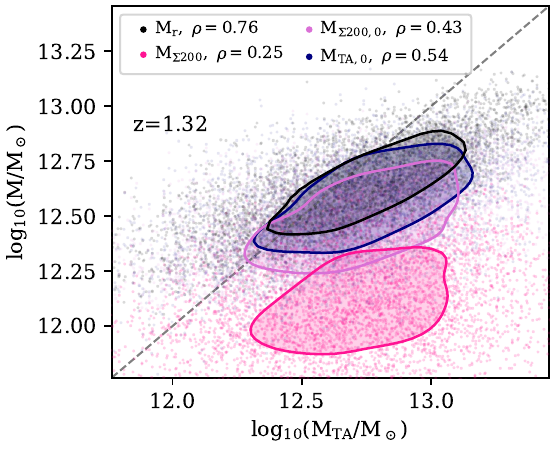} 
   \includegraphics[width=2.25in]{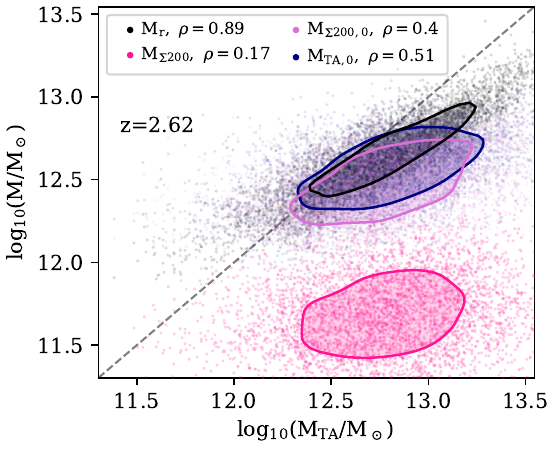} 
   \includegraphics[width=2.25in]{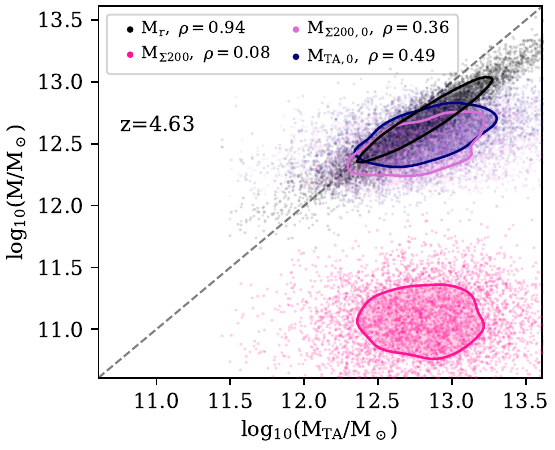}
   \vspace{-.1cm}
    \caption{Correlation between different mass estimators and $\mathrm{M_{TA}}$ at six different redshifts. Points show individual LG analogues from the `broad' sample, contours correspond to $1 \sigma$ of a Gaussian distribution. Black curves show $\mathrm{M_r}$, pink curves show $\mathrm{M_{\Sigma 200}}$. At $z>0$, we also show $\mathrm{M_{\Sigma 200, 0}}$ (in purple) and $\mathrm{M_{TA, 0}}$ (in blue). At every redshift, the correlation between $\mathrm{M_r}$ and $\mathrm{M_{TA}}$ is stronger than that between $\mathrm{M_{\Sigma 200}}$ and $\mathrm{M_{TA}}$, and the difference increases with increasing redshift. As expected, the correlation between $\mathrm{M_{TA,0}}$ and $\mathrm{M_{TA}}$ is strongest at low redshift, but also remains strong at higher redshifts. At all times, the correlation between $\mathrm{M_{\Sigma 200,0}}$ and $\mathrm{M_{TA}}$ is stronger than that between $\mathrm{M_{\Sigma 200}}$ and $\mathrm{M_{TA}}$, and the difference increases with increasing redshift.}
    \label{fig:correlations}
\end{figure*}

In the bottom row, we show the evolution of the sum of the progenitor masses, $\mathrm{M_{\Sigma 200}}$, the mass of the Timing Argument solution, $\mathrm{M_{TA}}$, and the ratio between the two, $\mathrm{M_{\Sigma 200}} / \mathrm{M_{TA}}$. The sum of the two halo masses, $\mathrm{M_{\Sigma 200}}$, increases by a factor of $\sim 30$, from $\sim 10^{11} \Ms$ at $z=4.6$ to $\sim 3 \times 10^{12} \Ms$ at $z=0$. Interestingly, the Timing Argument mass, $\mathrm{M_{TA}}$, shows much less evolution, with the median value decreasing by less than a factor of two over the same time. Consequently, the ratio between $\mathrm{M_{\Sigma 200}}$ and $\mathrm{M_{TA}}$ evolves significantly. It approaches unity at $z=0$, but is far below unity at earlier times.

\section{A time-invariant mass}\label{sec:time-invariant}
Previous works have noted a conceptual inconsistency in applying the Timing Argument to a system whose mass has, presumably, increased significantly over time \citep[e.g.][]{vanderMarel-2008, Yepes-2013}. However, we find that the LG analogues are, in fact, behaving close to what would be expected for systems whose mass remains nearly constant. This result is consistent with the analytical work of \citet{Penarrubia-2014}, who showed $\mathrm{M_{TA}}$ to be an adiabatic invariant.

This observation, combined with the fact  that the sum of the two halo masses is clearly not constant, leads us to look for a time-invariant mass associated with the Local Group that matches $\mathrm{M_{TA}}$. In Figure~\ref{fig:evolution-mass-estimators}, we compare the median evolution of the mass enclosed within spheres of different sizes, all centred on the midpoint of the MW-M31 system or their progenitors. On the left, we consider spheres of fixed physical sizes, in the middle, spheres of fixed comoving sizes, and on the right spheres whose radii are a multiple of the evolving separation of the MW-M31 pair.

The mass enclosed in a sphere of physical radius is constant if its density is constant. However, as shown in the left panel of Figure~\ref{fig:evolution-mass-estimators}, for all fixed radii, the enclosed masses evolve. For spheres of radius $R > 1000$ kpc, the density monotonically decreases, initially at a rate consistent with the expansion of the Universe, and then more slowly. For spheres of smaller radii, an initial period of decrease driven by expansion is followed by a period of increase due to the collapse of the LG system. While for some fixed physical radii, the enclosed mass varies considerably less than the $\mathrm{M_{\Sigma 200}}$ mass, no fixed-size sphere results in a constant mass.

A sphere fixed in comoving coordinates expands in line with the expansion of the Universe. The enclosed mass remains constant only if the evolution of the enclosed density matches that of the Universe. As shown in the middle panel of Figure~\ref{fig:evolution-mass-estimators}, this is not the case for our LG analogues out to any radius up to $R=2000$~$c$kpc. The LG analogues expand more slowly than the Universe on average, leading to an increase in the mass enclosed within comoving spheres.

In the right panel of Figure~\ref{fig:evolution-mass-estimators}, we show the mass enclosed within shells whose radius, $R$, is a multiple of the evolving separation, $r$, between the MW and M31 or their progenitors. For spheres of radius $R > r$, we find a monotonic decrease in mass over time. Here, the evolution is governed more by the expansion of the Universe than by the collapse of the LG analogue. At $R < r$, we find the opposite behaviour: the enclosed mass typically increases over time, now driven by the collapsing LG.

Interestingly, at $R \sim r$, the median enclosed mass is nearly constant, implying that there are no net in- or outflows through the surface of this sphere. Such spheres of radius $R = r$ thus provide a definition of the LG system that does not change with time. We subsequently refer to this enclosed mass, $\mathrm{M}(< R = r)$, as $\mathrm{M_r}$.

In Figure~\ref{fig:evolution-relative}, we compare the mass inside a sphere of radius equal to the separation, $\mathrm{M_r}$, to the mass inferred from the Timing Argument, $\mathrm{M_{TA}}$. Unsurprisingly, as both estimators are nearly constant in time, their ratio stays close to constant. More importantly, their ratio is also very close to unity: at all times, $\mathrm{M_{TA}}$ is close to $\mathrm{M_r}$. This is in stark contrast to the relation between $\mathrm{M_{TA}}$ and $\mathrm{M_{\Sigma 200}}$. While the sum of the two halo masses approaches the mass of the Timing Argument at $z=0$, it is far lower at earlier times. As indicated by the shaded areas, the $\pm 1 \sigma$ scatter in the relation between $\mathrm{M_{r}}$ and $\mathrm{M_{TA}}$ is also smaller than that in the relation between $\mathrm{M_{\Sigma 200}}$ and $\mathrm{M_{TA}}$.

In Figure~\ref{fig:correlations}, we show the correlations of $\mathrm{M_{\Sigma 200}}$ and $\mathrm{M_r}$ with $\mathrm{M_{TA}}$ for individual objects at six different redshifts. At $z > 0$, we additionally show the correlations of both $\mathrm{M_{\Sigma 200,0}}$ and $\mathrm{M_{TA,0}}$ with $\mathrm{M_{TA}}$. All correlation coefficients, $\rho$, are computed between the base-10 logarithms of the given quantities. Already at $z=0$, the correlation between $\mathrm{M_r}$ and $\mathrm{M_{TA}}$ is significantly stronger than that between $\mathrm{M_{\Sigma 200}}$ and $\mathrm{M_{TA}}$. This difference increases further with redshift: while the correlation between $\mathrm{M_{\Sigma 200}}$ and $\mathrm{M_{TA}}$ is very weak at redshifts above $z\gtrsim 1$, the correlation between $\mathrm{M_r}$ and $\mathrm{M_{TA}}$ is strong at all redshifts.

In fact, at $z > 1$, the correlation between $\mathrm{M_r}$ and $\mathrm{M_{TA}}$ becomes increasingly tight. This is explained by the fact that, as shown in Figure~\ref{fig:evolution-Timing-Argument} and consistent with the assumptions made in setting up Equations~\ref{eqn:TA-beginning}--\ref{eqn:TA-end}, the velocity component of the Timing Argument at high redshift is mostly driven by the universal expansion, while the lower density contrast means that the enclosed mass strongly correlates with the distance component. We limit our analysis to redshifts below $z \leq 4.6$ due to the mass resolution of our simulation, but see no reason why the method would not also extend to much higher redshifts.

Comparing the $\mathrm{M_{\Sigma 200,0}}$ and $\mathrm{M_{TA,0}}$ values with $\mathrm{M_{TA}}$, we find significant correlations at each redshift. As expected, the correlation between $\mathrm{M_{TA,0}}$ and $\mathrm{M_{TA}}$ is strongest close to $z=0$ but remains high at all times. Interestingly, at all redshifts above $z=0$, the correlation between $\mathrm{M_{\Sigma 200,0}}$ and $\mathrm{M_{TA}}$ is stronger than that between $\mathrm{M_{\Sigma 200}}$ and $\mathrm{M_{TA}}$. At any time $t < t_0$, $\mathrm{M_{TA}}$ is more strongly correlated with the sum of the halo masses at $z=0$ than with the sum of the halo masses at that time. This further illustrates that $\mathrm{M_{TA}}$ only indirectly depends on $\mathrm{M_{\Sigma 200}}$ via the total mass of the system, best captured by $\mathrm{M_r}$ and reasonably well approximated by $\mathrm{M_{\Sigma 200}}$ only close to $z=0$.

\section{Summary} \label{sec:summary}
In the context of the Local Group, the mass derived from the Timing Argumnt, $\mathrm{M_{TA}}$, has often been associated with the sum of the masses of the MW and M31. However, we show that, while $\mathrm{M_{TA}}$ is approximately constant over time, the sum of the halo masses is substantially lower for much of their evolution, and only approaches $\mathrm{M_{TA}}$ close to $z=0$. By contrast, we find that the Local Group mass that decouples from the expansion, that is constant over time, and  that is closely correlated with the Timing Argument value at all times, is contained within a sphere whose radius is equal to the separation of the pair. We identify this as the total mass of the Local Group.

Curiously, at times when both haloes contain only a few percent of their $z=0$ masses, the Timing Argument appears to already foretell their future growth. However, nothing in the Timing Argument marks out the present age of the Universe as special. Instead, this apparent premonition merely reflects the fact that the orbit always depends on the total mass of the system, of which the fraction contained within the two main haloes only approaches unity close to $z=0$.

At $z=0$, the mass within $r$ from the centre of our LG analogues is $25_{-8}^{+9} \%$ larger than the combined mass within $\mathrm{r_{200}}$ of the MW and M31. This alleviates the reported difference between the sum of the $\mathrm{M_{200}}$ values measured independently, and the mass inferred via the Timing Argument \citep{Benisty-2022, Sawala-2023}. It also supports the notion that dynamical measurements of the LG's mass, including via the virial theorem method \citep[e.g.][]{Diaz-2014} or Hubble flow perturbations \cite[e.g.][]{Penarrubia-2017}, measure a total mass that is greater than the sum of the masses of M31 and the MW.

In previous descriptions of the Timing Argument \citep[e.g.][]{Yepes-2013}, it has been stated that the assumption of constant point masses is a simplification made in the Timing Argument. We suggest that this is misguided: in fact, the growth of the two haloes is irrelevant to the Timing Argument. It is only the fact that, coincidentally, by $z=0$, the MW and M31 encompass a large fraction of the total mass that allows an (albeit not particularly accurate) conflation of the sum of their present masses with the mass of the system derived from the Timing Argument.

\section*{Data Availability Statement}
Our analysis is based entirely on public data available at http://skiesanduniverses.org. A script to produce all figures in the paper is available at https://github.com/TillSawala/TimingArgument.

\section*{Acknowledgements}
We thank the creators of the {\sc Uchuu} simulation for sharing their data, and AA-CSIC, CESGA and RedIRIS for hosting the Skies \& Universes service, supported by MICINN EU-Feder grant EQC2018-004366-P. We also thank the anonymous referee for their helpful suggestions.
TS and PHJ acknowledge support from Academy of Finland grant 339127. TS acknowledges support from European Research Council (ERC) Advanced Grant DMIDAS (GA 786910), and SL and PHJ acknowledge the support from the ERC Consolidator Grant KETJU (no. 818930). This work used facilities hosted by the CSC - IT Centre for Science, Finland, and the DiRAC@Durham facility managed by the ICC (www.dirac.ac.uk), with support from BEIS via STFC capital grants ST/K00042X/1, ST/P002293/1, ST/R002371/1 and ST/S002502/1, and STFC operations grant ST/R000832/1. We used open source software, including Matplotlib \citep{matplotlib-paper}, SciPy \citep{SciPy} and NumPy \citep{numpy-paper}.



\bibliographystyle{mnras} \bibliography{paper}

\begin{thebibliography}{}
\makeatletter
\relax
\def\mn@urlcharsother{\let\do\@makeother \do\$\do\&\do\#\do\^\do\_\do\%\do\~}
\def\mn@doi{\begingroup\mn@urlcharsother \@ifnextchar [ {\mn@doi@}
  {\mn@doi@[]}}
\def\mn@doi@[#1]#2{\def\@tempa{#1}\ifx\@tempa\@empty \href
  {http://dx.doi.org/#2} {doi:#2}\else \href {http://dx.doi.org/#2} {#1}\fi
  \endgroup}
\def\mn@eprint#1#2{\mn@eprint@#1:#2::\@nil}
\def\mn@eprint@arXiv#1{\href {http://arxiv.org/abs/#1} {{\tt arXiv:#1}}}
\def\mn@eprint@dblp#1{\href {http://dblp.uni-trier.de/rec/bibtex/#1.xml}
  {dblp:#1}}
\def\mn@eprint@#1:#2:#3:#4\@nil{\def\@tempa {#1}\def\@tempb {#2}\def\@tempc
  {#3}\ifx \@tempc \@empty \let \@tempc \@tempb \let \@tempb \@tempa \fi \ifx
  \@tempb \@empty \def\@tempb {arXiv}\fi \@ifundefined
  {mn@eprint@\@tempb}{\@tempb:\@tempc}{\expandafter \expandafter \csname
  mn@eprint@\@tempb\endcsname \expandafter{\@tempc}}}

\bibitem[\protect\citeauthoryear{{Behroozi}, {Wechsler}  \& {Wu}}{{Behroozi}
  et~al.}{2013a}]{Behroozi-2013a}
{Behroozi} P.~S.,  {Wechsler} R.~H.,   {Wu} H.-Y.,  2013a, \mn@doi [\apj]
  {10.1088/0004-637X/762/2/109}, \href
  {https://ui.adsabs.harvard.edu/abs/2013ApJ...762..109B} {762, 109}

\bibitem[\protect\citeauthoryear{{Behroozi}, {Wechsler}, {Wu}, {Busha},
  {Klypin}  \& {Primack}}{{Behroozi} et~al.}{2013b}]{Behroozi-2013b}
{Behroozi} P.~S.,  {Wechsler} R.~H.,  {Wu} H.-Y.,  {Busha} M.~T.,  {Klypin}
  A.~A.,   {Primack} J.~R.,  2013b, \mn@doi [\apj]
  {10.1088/0004-637X/763/1/18}, \href
  {https://ui.adsabs.harvard.edu/abs/2013ApJ...763...18B} {763, 18}

\bibitem[\protect\citeauthoryear{{Benisty}}{{Benisty}}{2021}]{Benisty-2021}
{Benisty} D.,  2021, \mn@doi [\aap] {10.1051/0004-6361/202142096}, \href
  {https://ui.adsabs.harvard.edu/abs/2021A&A...656A.129B} {656, A129}

\bibitem[\protect\citeauthoryear{{Benisty} \& {Capozziello}}{{Benisty} \&
  {Capozziello}}{2023}]{Benisty-2023b}
{Benisty} D.,  {Capozziello} S.,  2023, \mn@doi [Physics of the Dark Universe]
  {10.1016/j.dark.2023.101175}, \href
  {https://ui.adsabs.harvard.edu/abs/2023PDU....3901175B} {39, 101175}

\bibitem[\protect\citeauthoryear{{Benisty}, {Vasiliev}, {Evans}, {Davis},
  {Hartl}  \& {Strigari}}{{Benisty} et~al.}{2022}]{Benisty-2022}
{Benisty} D.,  {Vasiliev} E.,  {Evans} N.~W.,  {Davis} A.-C.,  {Hartl} O.~V.,
  {Strigari} L.~E.,  2022, \mn@doi [\apjl] {10.3847/2041-8213/ac5c42}, \href
  {https://ui.adsabs.harvard.edu/abs/2022ApJ...928L...5B} {928, L5}

\bibitem[\protect\citeauthoryear{{Benisty}, {Davis}  \& {Evans}}{{Benisty}
  et~al.}{2023}]{Benisty-2023}
{Benisty} D.,  {Davis} A.~C.,   {Evans} N.~W.,  2023, \mn@doi [arXiv e-prints]
  {10.48550/arXiv.2306.14963}, \href
  {https://ui.adsabs.harvard.edu/abs/2023arXiv230614963B} {p. arXiv:2306.14963}

\bibitem[\protect\citeauthoryear{{Chamberlain}, {Price-Whelan}, {Besla},
  {Cunningham}, {Garavito-Camargo}, {Pe{\~n}arrubia}  \&
  {Petersen}}{{Chamberlain} et~al.}{2023}]{Chamberlain-2023}
{Chamberlain} K.,  {Price-Whelan} A.~M.,  {Besla} G.,  {Cunningham} E.~C.,
  {Garavito-Camargo} N.,  {Pe{\~n}arrubia} J.,   {Petersen} M.~S.,  2023,
  \mn@doi [\apj] {10.3847/1538-4357/aca01f}, \href
  {https://ui.adsabs.harvard.edu/abs/2023ApJ...942...18C} {942, 18}

\bibitem[\protect\citeauthoryear{{Diaz}, {Koposov}, {Irwin}, {Belokurov}  \&
  {Evans}}{{Diaz} et~al.}{2014}]{Diaz-2014}
{Diaz} J.~D.,  {Koposov} S.~E.,  {Irwin} M.,  {Belokurov} V.,   {Evans} N.~W.,
  2014, \mn@doi [\mnras] {10.1093/mnras/stu1210}, \href
  {https://ui.adsabs.harvard.edu/abs/2014MNRAS.443.1688D} {443, 1688}

\bibitem[\protect\citeauthoryear{{Doroshkevich}}{{Doroshkevich}}{1970}]{Doroshkevich-1970}
{Doroshkevich} A.~G.,  1970, Astrofizika, \href
  {https://ui.adsabs.harvard.edu/abs/1970Afz.....6..581D} {6, 581}

\bibitem[\protect\citeauthoryear{{Guo} et~al.,}{{Guo} et~al.}{2011}]{Guo-2011}
{Guo} Q.,  et~al., 2011, \mn@doi [\mnras] {10.1111/j.1365-2966.2010.18114.x},
  \href {http://adsabs.harvard.edu/abs/2011MNRAS.413..101G} {413, 101}

\bibitem[\protect\citeauthoryear{Harris et~al.,}{Harris
  et~al.}{2020}]{numpy-paper}
Harris C.~R.,  et~al., 2020, \mn@doi [Nature] {10.1038/s41586-020-2649-2}, 585,
  357

\bibitem[\protect\citeauthoryear{{Hartl} \& {Strigari}}{{Hartl} \&
  {Strigari}}{2022}]{Hartl-2022}
{Hartl} O.~V.,  {Strigari} L.~E.,  2022, \mn@doi [\mnras]
  {10.1093/mnras/stac413}, \href
  {https://ui.adsabs.harvard.edu/abs/2022MNRAS.511.6193H} {511, 6193}

\bibitem[\protect\citeauthoryear{{Hoyle}}{{Hoyle}}{1951}]{Hoyle-1951}
{Hoyle} F.,  1951, in Problems of Cosmical Aerodynamics. p.~195

\bibitem[\protect\citeauthoryear{Hunter}{Hunter}{2007}]{matplotlib-paper}
Hunter J.~D.,  2007, \mn@doi [Computing in Science \& Engineering]
  {10.1109/MCSE.2007.55}, 9, 90

\bibitem[\protect\citeauthoryear{{Ishiyama} et~al.,}{{Ishiyama}
  et~al.}{2021}]{Ishiyama-2021}
{Ishiyama} T.,  et~al., 2021, \mn@doi [\mnras] {10.1093/mnras/stab1755}, \href
  {https://ui.adsabs.harvard.edu/abs/2021MNRAS.506.4210I} {506, 4210}

\bibitem[\protect\citeauthoryear{{Kahn} \& {Woltjer}}{{Kahn} \&
  {Woltjer}}{1959}]{Kahn-1959}
{Kahn} F.~D.,  {Woltjer} L.,  1959, \mn@doi [\apj] {10.1086/146762}, \href
  {https://ui.adsabs.harvard.edu/abs/1959ApJ...130..705K} {130, 705}

\bibitem[\protect\citeauthoryear{{Lemos}, {Jeffrey}, {Whiteway}, {Lahav},
  {Libeskind}  \& {Hoffman}}{{Lemos} et~al.}{2021}]{Lemos-2021}
{Lemos} P.,  {Jeffrey} N.,  {Whiteway} L.,  {Lahav} O.,  {Libeskind} N.,
  {Hoffman} Y.,  2021, \mn@doi [\prd] {10.1103/PhysRevD.103.023009}, \href
  {https://ui.adsabs.harvard.edu/abs/2021PhRvD.103b3009L} {103, 023009}

\bibitem[\protect\citeauthoryear{{Li} \& {White}}{{Li} \&
  {White}}{2008}]{Li-2008}
{Li} Y.-S.,  {White} S.~D.~M.,  2008, \mn@doi [\mnras]
  {10.1111/j.1365-2966.2007.12748.x}, \href
  {http://adsabs.harvard.edu/abs/2008MNRAS.384.1459L} {384, 1459}

\bibitem[\protect\citeauthoryear{{Lynden-Bell}}{{Lynden-Bell}}{1981}]{Lynden-Bell-1981}
{Lynden-Bell} D.,  1981, The Observatory, \href
  {https://ui.adsabs.harvard.edu/abs/1981Obs...101..111L} {101, 111}

\bibitem[\protect\citeauthoryear{{McLeod} \& {Lahav}}{{McLeod} \&
  {Lahav}}{2020}]{McLeod-2020}
{McLeod} M.,  {Lahav} O.,  2020, \mn@doi [\jcap]
  {10.1088/1475-7516/2020/09/056}, \href
  {https://ui.adsabs.harvard.edu/abs/2020JCAP...09..056M} {2020, 056}

\bibitem[\protect\citeauthoryear{{Partridge}, {Lahav}  \&
  {Hoffman}}{{Partridge} et~al.}{2013}]{Partridge-2013}
{Partridge} C.,  {Lahav} O.,   {Hoffman} Y.,  2013, \mn@doi [\mnras]
  {10.1093/mnrasl/slt109}, \href
  {https://ui.adsabs.harvard.edu/abs/2013MNRAS.436L..45P} {436, L45}

\bibitem[\protect\citeauthoryear{{Pe{\~n}arrubia}, {Ma}, {Walker}  \&
  {McConnachie}}{{Pe{\~n}arrubia} et~al.}{2014}]{Penarrubia-2014}
{Pe{\~n}arrubia} J.,  {Ma} Y.-Z.,  {Walker} M.~G.,   {McConnachie} A.,  2014,
  \mn@doi [\mnras] {10.1093/mnras/stu879}, \href
  {http://adsabs.harvard.edu/abs/2014MNRAS.443.2204P} {443, 2204}

\bibitem[\protect\citeauthoryear{{Peebles}}{{Peebles}}{1969}]{Peebles-1969}
{Peebles} P.~J.~E.,  1969, \mn@doi [\apj] {10.1086/149876}, \href
  {https://ui.adsabs.harvard.edu/abs/1969ApJ...155..393P} {155, 393}

\bibitem[\protect\citeauthoryear{Peñarrubia \& Fattahi}{Peñarrubia \&
  Fattahi}{2017}]{Penarrubia-2017}
Peñarrubia J.,  Fattahi A.,  2017, \mn@doi [Monthly Notices of the Royal
  Astronomical Society] {10.1093/mnras/stx323}, 468, 1300–1316

\bibitem[\protect\citeauthoryear{Peñarrubia, Gómez, Besla, Erkal  \&
  Ma}{Peñarrubia et~al.}{2015}]{Penarrubia-2015}
Peñarrubia J.,  Gómez F.~A.,  Besla G.,  Erkal D.,   Ma Y.-Z.,  2015, \mn@doi
  [Monthly Notices of the Royal Astronomical Society: Letters]
  {10.1093/mnrasl/slv160}, 456, L54–L58

\bibitem[\protect\citeauthoryear{{Planck Collaboration} et~al.,}{{Planck
  Collaboration} et~al.}{2020}]{Planck-2020}
{Planck Collaboration} et~al., 2020, \mn@doi [\aap]
  {10.1051/0004-6361/201935891}, \href
  {https://ui.adsabs.harvard.edu/abs/2020A&A...641A...9P} {641, A9}

\bibitem[\protect\citeauthoryear{{Savitzky} \& {Golay}}{{Savitzky} \&
  {Golay}}{1964}]{Savitzky-1964}
{Savitzky} A.,  {Golay} M.~J.~E.,  1964, \mn@doi [Analytical Chemistry]
  {10.1021/ac60214a047}, \href
  {https://ui.adsabs.harvard.edu/abs/1964AnaCh..36.1627S} {36, 1627}

\bibitem[\protect\citeauthoryear{{Sawala}, {Teeriaho}  \& {Johansson}}{{Sawala}
  et~al.}{2023}]{Sawala-2023}
{Sawala} T.,  {Teeriaho} M.,   {Johansson} P.~H.,  2023, \mn@doi [\mnras]
  {10.1093/mnras/stad883}, \href
  {https://ui.adsabs.harvard.edu/abs/2023MNRAS.521.4863S} {521, 4863}

\bibitem[\protect\citeauthoryear{Virtanen et~al.,}{Virtanen
  et~al.}{2020}]{SciPy}
Virtanen P.,  et~al., 2020, \mn@doi [Nature Methods]
  {10.1038/s41592-019-0686-2}, \href {https://rdcu.be/b08Wh} {17, 261}

\bibitem[\protect\citeauthoryear{{White}}{{White}}{1984}]{White-1984}
{White} S.~D.~M.,  1984, \mn@doi [\apj] {10.1086/162573}, \href
  {https://ui.adsabs.harvard.edu/abs/1984ApJ...286...38W} {286, 38}

\bibitem[\protect\citeauthoryear{{White} \& {Frenk}}{{White} \&
  {Frenk}}{1991}]{White-1991}
{White} S. D.~M.,  {Frenk} C.~S.,  1991, \mn@doi [\apj] {10.1086/170483}, \href
  {https://ui.adsabs.harvard.edu/abs/1991ApJ...379...52W} {379, 52}

\bibitem[\protect\citeauthoryear{{White} \& {Rees}}{{White} \&
  {Rees}}{1978}]{White-1987}
{White} S.~D.~M.,  {Rees} M.~J.,  1978, \mn@doi [\mnras]
  {10.1093/mnras/183.3.341}, \href
  {https://ui.adsabs.harvard.edu/abs/1978MNRAS.183..341W} {183, 341}

\bibitem[\protect\citeauthoryear{{Yepes}, {Gottl{\"o}ber}  \&
  {Hoffman}}{{Yepes} et~al.}{2014}]{Yepes-2013}
{Yepes} G.,  {Gottl{\"o}ber} S.,   {Hoffman} Y.,  2014, \mn@doi [New Astronomy
  Reviews] {10.1016/j.newar.2013.11.001}, \href
  {http://adsabs.harvard.edu/abs/2014NewAR..58....1Y} {58, 1}

\bibitem[\protect\citeauthoryear{van~der Marel \& Guhathakurta}{van~der Marel
  \& Guhathakurta}{2008}]{vanderMarel-2008}
van~der Marel R.~P.,  Guhathakurta P.,  2008, \mn@doi [The Astrophysical
  Journal] {10.1086/533430}, 678, 187–199

\bibitem[\protect\citeauthoryear{{van der Marel}, {Fardal}, {Besla}, {Beaton},
  {Sohn}, {Anderson}, {Brown}  \& {Guhathakurta}}{{van der Marel}
  et~al.}{2012}]{vanderMarel-2012}
{van der Marel} R.~P.,  {Fardal} M.,  {Besla} G.,  {Beaton} R.~L.,  {Sohn}
  S.~T.,  {Anderson} J.,  {Brown} T.,   {Guhathakurta} P.,  2012, \mn@doi
  [\apj] {10.1088/0004-637X/753/1/8}, \href
  {http://adsabs.harvard.edu/abs/2012ApJ...753....8V} {753, 8}

\makeatother
\end{thebibliography}


\bsp	
\label{lastpage}
\end{document}